\documentstyle[prl,multicol,aps,psfig]{revtex}
\newcommand{\be}{\begin{equation}}
\newcommand{\ee}{\end{equation}}
\newcommand{\PP}{{\mathcal{P}}}
\newcommand{\HH}{{\mathcal{H}}}

\begin{document}
\title{Exchange Frequencies in the 2d Wigner crystal }
\author{B. Bernu$^1$, Ladir C\^andido$^2$ and  D. M. Ceperley$^2$}
\address{$^1$Laboratoire de Physique Th\'eorique des Liquides, UMR 7600
of CNRS, Universit\'e P. et M. Curie, boite 121, 4 Place Jussieu,
75252 Paris, France}
 \address{$^2$Dept. of Physics and NCSA University of Illinois at
Urbana-Champaign, Urbana, IL 61801}
 \maketitle

\begin{abstract}
Using Path Integral Monte Carlo we have calculated exchange
frequencies as electrons undergo ring exchanges in a ``clean'' 2d
Wigner crystal as a function of density. The results show
agreement with WKB calculations at very low density, but show a
more rapid increase with density near melting. Remarkably, the
exchange Hamiltonian closely resembles the measured exchanges in
2d $^3$He. Using the resulting multi-spin exchange model we find
the spin Hamiltonian for $r_s \leq 175 \pm 10 $ is a frustrated
antiferromagnetic; its likely ground state is a spin liquid. For
lower density the ground state will be ferromagnetic.
\end{abstract}

\vspace*{3mm} \noindent PACS Numbers:  73.20Dx, 75.10-b, 67.8Jd


\begin{multicols}{2}
\narrowtext


The uniform system of electrons is one of the basic models of
condensed matter physics. In this paper, we report on the first
exact calculations of the spin Hamiltonian in the low density 2
dimensional Wigner crystal (2dWC) near melting. This system is
realized experimentally with electrons confined at an
semiconductor MOSFET's and heterostructures\cite{yoon}, and for
electrons on the surface of liquid helium\cite{grimes}.

A homogeneous charged system is characterized by two parameters:
the density given in terms of $r_s = a/a_0 =( m^*/m\epsilon )(\pi
a_0^2 \rho)^{-1/2}$ and the energy in effective Rydbergs
$Ry^*=(m^*/m_e \epsilon^2)Ry$ where $m^*$ is the effective mass
and $\epsilon$ the dielectric constant. 
Figure 1 summarizes the 2d phase diagram. At low density (large
$r_s$) the potential energy dominates over the kinetic energy and
the system forms a perfect triangular lattice, the Wigner
crystal\cite{zhu}. Tanatar and Ceperley\cite{tanatar} determined
that melting at zero temperature occurs at $r_s \simeq 37 \pm 5$.
Recent calculations\cite{vac} have shown that the low temperature
phase is free of point defects for densities with $r_s \geq 50$
but defects may be present very near melting. At densities for
$r_s \geq 100$ the melting is classical, and occurs for
temperatures $T_{melt} = 2Ry /(\Gamma_c r_s)$ where $\Gamma_c
\approx 137$\cite{strandburg}.

We determine the spin-spin interaction in the Wigner crystal,
using Thouless'\cite{thouless} theory of exchange. According to
this theory, in the absence of point defects, at low temperatures
the spins will be governed by a Hamiltonian of the form:
 \be
\HH_{spin} = - \sum_P (-1)^P J_P {\hat \PP}_{spin}
 \ee
where the sum is over all cyclic (ring) exchanges described by a
cyclic permutation $P$, $J_P$ is its exchange frequency and ${\hat
\PP}_{spin}$ is the corresponding spin exchange operator. Path
Integral Monte Carlo (PIMC) as suggested by
Thouless\cite{thouless} and Roger\cite{roger84}has proved to be
the only reliable way to calculate these parameters. The theory
and computational method have been tested thoroughly on the
magnetic properties of bulk helium obtaining excellent agreement
with measured properties\cite{RMPI}. Rather surprisingly, it has
been found\cite{RHD} that in both 2d and 3d solid $^3$He,
exchanges of 2, 3 and 4 particles have roughly the same order of
magnitude and must all be taken into account. This is known as the
multiple spin exchange model(MSE).

\begin{figure}
\centerline{\psfig{figure=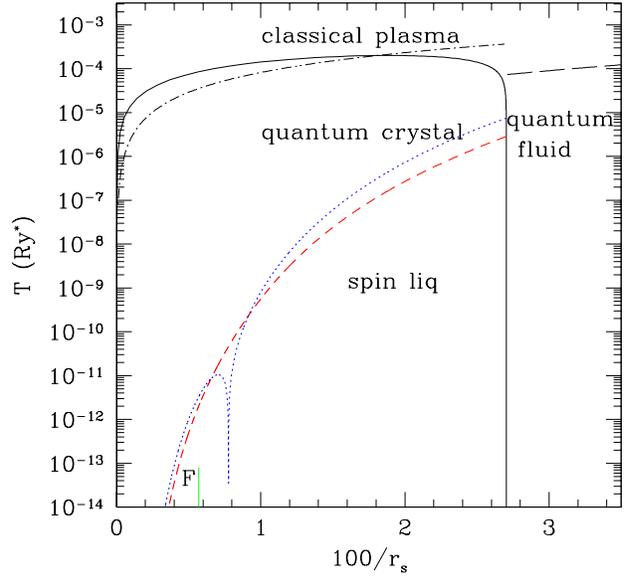,width=10cm,angle=0}}
\caption{Phase diagram. The estimated melting line is based on
Lindemann's criteria\protect\cite{douchin}. The long-dashed line
represents the Debye temperature. The dotted line is $\theta$, the
short dashed line is $J_c$. The vertical line is the estimated
zero temperature ferromagnetic (F) transition.}
\end{figure}

A WKB calculation of the exchange frequencies in the 2dWC by
Roger\cite{roger84} predicted that the three electron $J_3$
nearest neighbor exchange would dominate, leading to a
ferromagnetic(F) ground state. It has been recently pointed
out\cite{CKNV} that if the 2dWC can be stabilized by disorder to
higher densities, a transition to an antiferromagnetic(AF) phase
will occur. Although the 1/r interaction is characterized as
'soft', in the 2dWC the two particle pair correlation function for
the electron system is quite similar to that of solid $^3$He
supporting the idea that multiple exchanges could be important in
the 2dWC near melting.

The method of determining the exchange frequency in quantum
crystals has been given earlier\cite{RMPI,CJ,cornell}. One
computes the free energy necessary to make an exchange beginning
with one arrangement of particles to lattice sites $Z$ and ending
on a permuted arrangement $PZ$:
 \be F_P(\beta) =
Q(P,\beta)/Q(I,\beta) = \tanh(J_P(\beta-\beta_0)).
 \ee
Here $Q(P,\beta)$ is the partition function corresponding to an
exchange $P$ at a temperature $1/\beta$. $I$ is the identity
permutation. Note that these paths are of ``distinguishable''
particles since Fermi statistics are implemented through the spin
Hamilitonian in Eq. (1). We determine the function $F_P (\beta)$
using a method which directly calculates free energy differences
and thereby determines $J_P$ and $\beta_0$. The only new feature
with respect to the $^3$He calculations is the method of treating
the long-range potential. We use the standard Ewald
breakup\cite{roger84} and treat the short-range part using the
exact pair action\cite{RMPI} and the long-range k-space term using
the primitive approximation. We used a hexagonal unit cell with
periodic boundary conditions, most calculations containing 36
electrons. Checks with up to 144 electrons did not change the
results within the statistical errors. The number of particles is
not as important in the 2dWC as in solid $^3$He because the $1/r$
interaction suppresses the long wavelength charge fluctuations.

\begin{figure}
\centerline{\psfig{figure=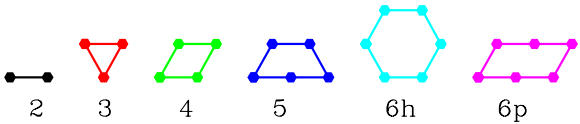,width=3in,height=1in
,bbllx=2in,bblly=8in,bburx=8in,bbury=10in }}
 \caption{Considered exchanges.} \label{diagram}
\end{figure}

We performed calculations for the densities $r_s = \{45,
50,60,75,100,140,200\}$.  We found accurate results using a
``time-step'' for the descretized imaginary time path integrals of
$\tau \leq 0.3 r_s^{3/2}$ and extrapolated to the $\tau=0$ limit
using $J_p(\tau)= J_p(0) + J_p' \tau^3$. The inverse temperature
$\beta$ in Eq. (2 )must be larger than the values determined by
melting and twice the exchange ``time'' $\beta_0 \sim  5
r_s^{3/2}$. We have determined the exchange frequencies with an
accuracy between 1.5\% and 6\% independent of their magnitude or
the number of exchanging electrons though the computer time
increases with the number of exchanging electrons. Breakdown of
Thouless' theory caused by many states contributing to the ratio
of partition functions would be signaled by $F_P (\beta)$ not
described by Eq. (2). Except for $r_s < 50$ where our calculations
are too unstable to make definite predictions, we observed no
problems of convergence.


Figure 2 shows the ring exchanges considered here. Except near
melting these exchanges give rise to most of the thermodynamic
properties. Note that we consider the 6-particle parallelogram
(6p) exchange, which is not taken into account in solid $^3$He. We
have also calculated several 2-5 particle exchanges having
next-nearest neighbor exchanges and all possible 6 particle
nearest-neighbor exchanges for $50 \leq r_s \leq 75$, but because
their magnitudes are much smaller, we do not report those results.

\begin{figure}
\centerline{\psfig{figure=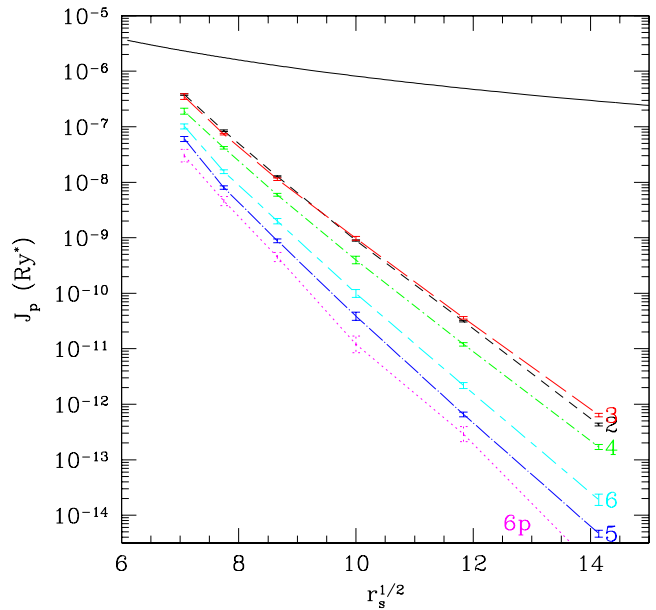,width=10cm,angle=00}} \caption{
Exchange frequencies versus $r_s^{1/2}$. The solid line is
$10^{-3}$ of the kinetic energy. }
\end{figure}

Our calculated exchange frequencies vary rapidly with density as
shown in fig. (3). One can see that they are much less than the
zero point energy of the electrons, thus justifying the use of
Thouless' theory. The WKB method\cite{RHD} where one approximates
the Path Integral in Eq. (2) by the single most probable path,
explains most of this density dependence.  In the 2dWC, the WKB
expression for the exchange frequency\cite{roger84,CKNV} is:
\be
J_P = A_P (r_s) b_P^{1/2} r_s^{-5/4} e^{-b_P r_s^{1/2} }.
 \ee
Here $b_P r_s^{1/2}$ is the minimum value of the action integral
along the path connecting $PZ$ with $Z$. The 3 particle exchange
exponent is the smallest indicating that as
$r_s\rightarrow\infty$, $J_3$ will dominate and the system will
have a ferromagnetic ground state. However, note that in fig. (3)
$J_2 > J_3$ for $r_s \leq 90$.  We observe a more rapid increase
in all the exchange frequencies than predicted by Eq. (3) for $r_s
<100$ caused by non-linear fluctuations about the classical path.
Near melting the spread in exchange frequencies is much less than
in the low density limit.

\begin{figure}
\centerline{\psfig{figure=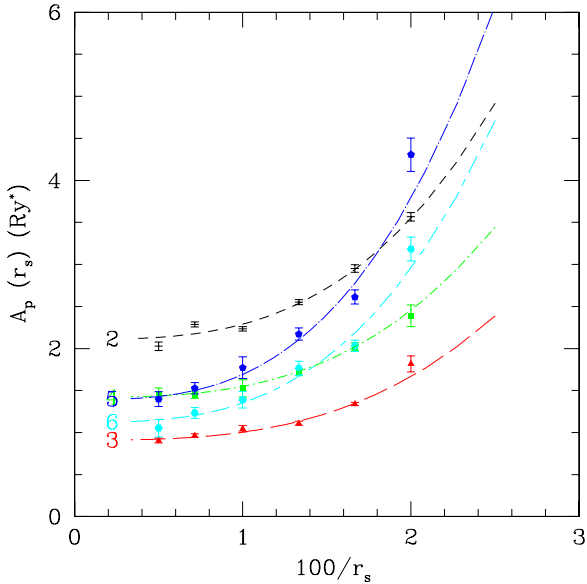,width=10cm,angle=00}}
\caption{The prefactor of the exchange frequencies, $A_P$ , see
Eq. (3) versus density. All prefactors go to a constant for
$r_s>100$ indicating that exchange frequencies are dominated by a
single path (WKB limit). The curves are fits as described in the
text. }
\end{figure}

Figure (4) shows density dependence of the prefactor, $A_P$. The
prefactors are larger than unity as expected\cite{CKNV} (our
definition of $A_p$ differs from ref. \cite{CKNV} by a factor of
$1.63/\sqrt{2\pi}$). We observe that the prefactor is roughly
constant for $r_s >100$ but rises rapidly as the system approaches
the melting density. We are unaware of any work that definitively
establishes the form of the correction to the WKB formula. A good
fit was obtained with the function $A_P (r_s) = A_P(0)
(1+(r_P/r_s)^3)$ to determine $b_P$, $A_P (0)$ and $r_P$. The
exponents are close to the WKB values, only differing
significantly for the 2-particle exchange.

The prefactor for $J_2$ is twice as large as the other exchanges.
One might think that since one cannot observe the difference in
pair exchanges in the initial and final spins, the sense of the
exchange would not matter. Our results show that the 2
orientations of 2 particle exchange, clockwise and
counter-clockwise, should be counted separately. A recent
calculation\cite{flambaum} directly evaluates $J_2$ in the high
density regime, $5\leq r_s \leq 45$, by numerically solving for
the difference between the even and odd parity 2-electron
energies. In those calculations, the spectator electrons were
fixed at their lattice sites and the 2 exchanging electrons had
vanishing wavefunctions outside a rectangle centered around the
exchange. At $r_s =45$ a direct comparison shows their result is
2.3 times larger than that obtained with PIMC. This result is
expected because of the additional localization caused by the
spectator electrons fluctuating into the exchanging region.


Having determined the exchange frequency, one is left with the
spin Hamiltonian of Eq. (1). This is a non-trivial many-body
problem which we will not discuss in detail here. For spin 1/2
systems, $J_2$ and $J_3$ contribute only with a nearest neighbor
Heisenberg term: $J_2^{\rm eff} = J_2 -2 J_3$. This term is
negative (ferromagnetic) but approaches zero near melting. For
convenience we use $J_4$ as a reference to fix the overall scale
of the magnetic energy. Neglecting $J_{6p}$, the Hamiltonian has 3
remaining parameters $J_2^{\rm eff}/J_4$, $J_5/J_4$ and
$J_{6h}/J_4$. The dependence of these ratios on density is shown
in fig. (5).

\begin{figure}
\centerline{\psfig{figure=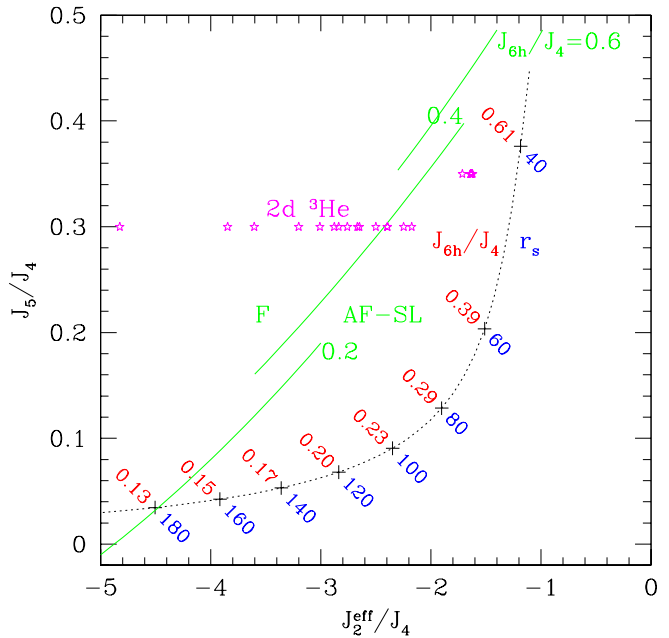,width=10cm,angle=0}}
\caption{Spin Phase diagram as a function of exchange ratios. The
dotted line is the flow of spin hamiltonian space versus $r_s$
(lower numbers); also shown the are estimated values of
$J_{6h}/J_4$(upper numbers).  The solid lines are the limit of the
ferromagnetic phase according to ED\protect\cite{misguich} at
$J_{6h}/J_4= \{0.2, 0.4, 0.6\}$. The 2dWC crosses into the F
region for $r_s\approx 175$. The $(\star)$ are empirical estimates
of the spin Hamiltonian of 2d $^3$He at several
densities\protect\cite{roger98}. }
\end{figure}

High temperature series expansions \cite{roger97}  determine the
the specific heat $C_V$ and magnetic susceptibility $\chi_0/\chi$
for temperatures $k_B T \gg J_P$.  The susceptibility is given by:
$\chi_0/\chi = T - \theta +B/T \ldots$ and the specific heat
$C_V/Nk_B = (3J_c/2T)^2 + \ldots$  where the Curie-Weiss constant
is given by $\theta = -3(J_2^{\rm eff}+3J_4-5 J_5 +5/8 J_{6h}+15/4
J_{6p} )$ with a quadratic expression of the $J$'s for $J_c$.
These two constants which set the scale of the temperature where
exchange is important are shown as a dotted and dashed lines on
fig. (1).
Both $\theta$ and $J_c$ decrease very rapidly at low density
showing that experiments must be done at $r_s \leq 100$ if spin
effects are to be at a reasonable temperature, e.g. $T_c > 0.1 mK$
(assuming $\epsilon=m^*=1$). Note that $\theta$ changes from
positive to negative at $r_s \approx 130$.

The zero temperature state can be studied by exact diagonalization
(ED) of an $N$ site system. (The present limitation is $N\lesssim
36$.) Here we focus on the zero temperature transition from a
ferromagnetic to an antiferromagnetic state. An estimate
\cite{misguich} of the  critical transition parameters is shown on
fig.(5). The ferromagnetic phase is obtained only at very low
density: we estimate the the F-AF transition for the 2dWC will
occur at $r_s =175 \pm 10$. (Note that this estimate does not
include the effect of $J_{6p}$; this will increase the stability
of the antiferromagnetic region to roughly $r_s=200$.)

At higher density, the frustration between large cycle exchanges
(4-6 body loops) results in a disordered spin state. For example,
the point ($J_2^{\rm eff}/J_4=-2$, $J_5=0$, $J_{6h}=0$), close to
the parameters at $r_s=100$, is a spin liquid with a gap to all
excitations\cite{misguich}: The spin liquid properties can be
understood from a resonance valance bond model with ordering into
spin-1 diamond plaquettes. The breakup of the plaquettes is
responsible for a low temperature peak on the specific heat. A
second high temperature peak develops at a temperature $T \sim
J_c$, shown as a dotted line on fig. (1).

We find that at higher densities ($r_s<100$), the trajectory of
the MSE models parallels the F-AF phase line, with the possibility
of a re-entrant ferromagnetic phase for $r_s<40$. Note that QMC
calculations\cite{varsano} of the normal fermi liquid at $r_s=30$
show that the ferromagnetic phase has a slightly lower energy that
the unpolarized phase. Hence both the high density 2dWC and the
low density electron fluid are characterized by a spin Hamiltonian
which is nearly ferromagnetic.

We note a remarkable similarity between the exchange parameters of
the 2dWC to those extrapolated from measurements of the second
layer of $^3$He absorbed on grafoil\cite{roger98} as shown in fig.
(5). The existence of these two related spin systems should allow
a fuller understanding of this remarkable spin liquid. Such an
equivalence could arise from  an underlying virtual
vacancy-interstitial (VI) mechanism\cite{RHD} giving rise to ring
exchanges. In this model the prefactor of the exchange is
controlled by the rate of VI formation (non-universal) but the
ratios of the various ring exchanges arises from geometry of the
triangular lattice and from the attraction of point defects
(universal). We have recently determined\cite{vac} that the VI
formation energy vanishes at melting in the 2dWC. This is
consistent with the fact that the various $J_P$ increase rapidly
near melting.

In summary, we find that the magnetic interactions are
characterized by a frustrated spin order. Application of a
magnetic field\cite{OK} transforms the exchange frequencies to
$J_p e^{ 2 \pi i e B_{\bot} a_p/h}$ where $a_p$ is the area of the
exchange (see table I) and $B_{\bot}$ the magnetic field.
Experiments with magnetic fields will allow exploration of this
Aharonov-Bohm effect and thereby provide direct information on
ring exchanges.

The semiconductor realizations of the 2dWC have significant
disorder which can stabilize localized electronic states at higher
densities than in the clean system\cite{chui}. One can also
stabilize the Wigner crystal at higher densities using bilayers
\cite{senatore}. Exchange frequencies in those systems including
effects of layer thickness and the exchange properties of point
defects could be calculated with the present method.

This research was funded by NSF DMR 98-02373, the CNRS-University
of Illinois exchange agreement, support by Funda\c c\~ ao  de
Amparo \`a Pesquisa do Estado de S\~ao Paulo (FAPESP) and the
Dept. of Physics at the University of Illinois. We used the
computational facilities at the NCSA.
\begin{table}
\begin{tabular}{|c|c|c|c|c|}
 P   & $b_P$ & $A_P(0)$& $r_P$ & $a_P$ \\\hline
  2  & 1.612       &  2.11  & 44 &           \\
  3  & 1.525       &  0.91  & 47 & 2.07(1)   \\
  4  & 1.656       &  1.42  & 45 & 3.07(2)   \\
  5  & 1.912       &  1.39  & 60 & 4.11(2)   \\
  6h & 1.790       &  1.12  & 59 & 7.04(3)   \\
  6p & 2.136       &  6.24  & 38 & 5.09(2)   \\
\end{tabular}
\caption{ Fits for the exchange frequencies.  The path area,
$a_P$, was calculated at $r_s=60$ and it is in units of triangle
areas. }
\end{table}

\end{multicols}
\widetext \vspace*{-5mm}
\begin{multicols}{2}

\end{multicols}

\end{document}